\documentclass[nofootinbib,prd,twocolumn,showpacs,showkeys,preprintnumbers]{revtex4-1}
\usepackage{hyperref,amssymb,amsmath,mathrsfs,bm,graphicx}
\begin{document}
\title {Quasi--homologous evolution of self--gravitating systems with vanishing complexity factor}
\author{L. Herrera}
\email{lherrera@usal.es}
\affiliation{Instituto Universitario de F\'isica
Fundamental y Matem\'aticas, Universidad de Salamanca, Salamanca 37007, Spain}
\author{A. Di Prisco}
\email{alicia.diprisco@ciens.ucv.ve}
\affiliation{Escuela de F\'\i sica, Facultad de Ciencias, Universidad Central de Venezuela, Caracas 1050, Venezuela}
\author{J. Ospino}
\email{j.ospino@usal.es}
\affiliation{Departamento de Matem\'atica Aplicada and Instituto Universitario de F\'isica
Fundamental y Matematicas, Universidad de Salamanca, Salamanca 37007, Spain}

\date{\today}
\begin{abstract}
We investigate the evolution of self--gravitating either dissipative or non--dissipative systems satisfying  the condition of minimal complexity, and whose areal radius velocity is proportional to the areal radius (quasi--homologous condition). Several exact analytical models are found under the above mentioned conditions. Some of the presented models   describe the evolution of spherically symmetric dissipative fluid distributions  whose center is surrounded by a cavity. Some of them  satisfy the Darmois conditions whereas others present shells and must satisfy the Israel condition on either one or both boundary surfaces. Prospective applications of some of these models to astrophysical scenarios are discussed.
\end{abstract}
\date{\today}

\maketitle
\section{Introduction}
\label{intro}
In a recent paper \cite{1} a  concept aiming to asses the degree of complexity of a self--gravitating spherically symmetric static fluid distribution was introduced  with the hope that the variable defining such property could help to deepen in the study of  self--gravitating systems. This definition was latter on extended to the time dependent case \cite{2}, which required the introduction of  a criterium for a definition of the simplest  pattern of evolution. The presented arguments in \cite{2} strongly suggested that the homologous condition  seemed to be the most suitable to describe the simplest  mode of evolution.  The  applications of this concept including systems with different kind of symmetry and/or other theories of gravity, may be found in \cite{3,4,5,6,7,8,9,10,11,12,13,14com,15com,16com,17com,18com,19com,20com,21com} and references therein.

In this work we are concerned with the problem of general relativistic gravitational collapse under the assumption of vanishing complexity factor.   The relevance of the study of gravitational collapse in astrophysics is illustrated by the fact, that the gravitational collapse of massive stars represents one of
the few observable phenomena where general relativity is expected to
play a relevant role.  Ever since the early work by Oppenheimer and Snyder \cite{Opp}, much has been done  by
researchers trying to provide models of evolving self--gravitating spheres.
However this endeavour proved to be difficult and uncertain. 

Thus, while it is true that numerical methods   enable researchers to investigate systems which are extremely
difficult to handle analytically, it is also true  that  purely
numerical solutions usually hinder to catch general, qualitative, aspects of the process.

On the other hand, analytical solutions although more suitable for a
general discussion, are sometimes found, either for too simplistic equations of state and/or under additional
heuristic assumptions whose justification is usually uncertain.

Notwithstanding we shall deal here with  analytical solutions  which are simple to analyze but still contain some of the essential features of a realistic situation. Different  methods for finding such solutions and their analysis have been  presented by many authors in recent years  (see \cite{relax,relaxII,relax3,relax4,relax5,relax6,sol1,sol2,sol3,sol4,sol5,sol6,sol7,sol8,sol9,sol10,sol11} and references therein).

In this work we endeavor to find exact solutions describing  either  dissipative or non--dissipative  fluid spheres satisfying the condition of vanishing  complexity factor, however in what concerns  the condition on the pattern of evolution, we shall relax the homologous condition assumed in \cite{2}  and will assume a much less stringent condition which will be referred to as  the quasi--homologous regime. The motivation to undertake this task is threefold. First,  it is of  interest to find the physical properties inherent to all dissipative systems characterized by a vanishing complexity factor. On the other hand, as we discovered during this research work,   the homologous condition appears to be too stringent, ruling out thereby many interesting scenarios from the astrophysical point of view.  Finally, the assumed ansatz provide a general method  for the obtention of analytical solutions to the Einstein equations, describing evolving fluid distributions.

Thus, the obtained models follow from the two conditions mentioned above plus some additional restrictions  on  kinematical variables. Some of the presented models describe fluid distributions with a vacuum cavity surrounding the center of symmetry  whereas others describe fluid distributions  filling the whole object. In the former case matching conditions at  both delimiting surfaces have to be  considered. The physical properties of all these models will be analyzed in detail and their eventual application to different astrophysical scenarios will be discussed.
\section{THE GENERAL SETUP OF THE PROBLEM: NOTATION, VARIABLES AND EQUATIONS}
\label{sec:1}

We consider spherically symmetric distributions  of collapsing fluid, which are  bounded by a spherical surface $\Sigma^{(e)}$, and, in the case that  a cavity  is present, are also bounded from inside by a spherical surface   $\Sigma^{(i)}$. The fluid is
assumed to be locally anisotropic (principal stresses unequal) and undergoing dissipation in the
form of heat flow (diffusion approximation). 

Thus the general line element may be written as
\begin{equation}
ds^2=-A^2dt^2+B^2dr^2+R^2(d\theta^2+\sin^2\theta d\phi^2),
\label{1}
\end{equation}
where the functions $A, B, R$ depend on $t$ and $r$.

The energy--momentum tensor  takes the form
\begin{eqnarray}
T_{\alpha\beta}&=&(\mu +
P_{\perp})V_{\alpha}V_{\beta}+P_{\perp}g_{\alpha\beta}+(P_r-P_{\perp})\chi_{
\alpha}\chi_{\beta}\nonumber \\&+&q_{\alpha}V_{\beta}+V_{\alpha}q_{\beta}
, \label{3}
\end{eqnarray}
where $\mu$ is the energy density, $P_r$ the radial pressure,
$P_{\perp}$ the tangential pressure, $q^{\alpha}$ the heat flux, $V^{\alpha}$ the four velocity of the fluid,
and $\chi^{\alpha}$ a unit four vector along the radial direction. These quantities
satisfy
\begin{eqnarray}
V^{\alpha}V_{\alpha}=-1, \;\; V^{\alpha}q_{\alpha}=0, \;\; \chi^{\alpha}\chi_{\alpha}=1,\;\;
\chi^{\alpha}V_{\alpha}=0.
\end{eqnarray}

It will be convenient to express the  energy momentum tensor  (\ref{3})  in the equivalent (canonical) form
\begin{equation}
T_{\alpha \beta} = {\mu} V_\alpha V_\beta + P h_{\alpha \beta} + \Pi_{\alpha \beta} +
q \left(V_\alpha \chi_\beta + \chi_\alpha V_\beta\right), \label{Tab}
\end{equation}
with
$$ P=\frac{P_{r}+2P_{\bot}}{3}, \qquad h_{\alpha \beta}=g_{\alpha \beta}+V_\alpha V_\beta,$$

$$\Pi_{\alpha \beta}=\Pi\left(\chi_\alpha \chi_\beta - \frac{1}{3} h_{\alpha \beta}\right), \qquad \Pi=P_{r}-P_{\bot}.$$

Since we are considering comoving observers, we have
\begin{eqnarray}
V^{\alpha}&=&A^{-1}\delta_0^{\alpha}, \;\;
q^{\alpha}=qB^{-1}\delta^{\alpha}_1, \;\;
\chi^{\alpha}=B^{-1}\delta^{\alpha}_1. \label{k}
\end{eqnarray}

It is worth noticing that we do not add explicitly  bulk or shear viscosity to the system because they
can be trivially absorbed into the radial and tangential pressures, $P_r$ and
$P_{\perp}$, of the collapsing fluid (in $\Pi$). Also we do not explicitly  introduce  dissipation in the free streaming approximation since it can be absorbed in $\mu, P_r$ and $q$. 
\subsection{Einstein equations}
\label{sec:2}
Einstein's field equations for the interior spacetime (\ref{1}) are given by
\begin{equation}
G_{\alpha\beta}=8\pi T_{\alpha\beta}.
\label{2}
\end{equation}
The non null components of (\ref{2})
with (\ref{1}) and  (\ref{3}),
read
\begin{widetext}
\begin{eqnarray}
8\pi T_{00}=8\pi  \mu A^2
=\left(2\frac{\dot{B}}{B}+\frac{\dot{R}}{R}\right)\frac{\dot{R}}{R}
-\left(\frac{A}{B}\right)^2\left[2\frac{R^{\prime\prime}}{R}+\left(\frac{R^{\prime}}{R}\right)^2
-2\frac{B^{\prime}}{B}\frac{R^{\prime}}{R}-\left(\frac{B}{R}\right)^2\right],
\label{T00} 
\end{eqnarray}
\end{widetext}
\begin{eqnarray}
8\pi T_{01}=-8\pi  q AB=-2\left(\frac{{\dot
R}^{\prime}}{R}-\frac{\dot B}{B}\frac{R^{\prime}}{R}-\frac{\dot
R}{R}\frac{A^{\prime}}{A}\right),
\label{T01} 
\end{eqnarray}
\begin{widetext}
\begin{eqnarray}
8\pi T_{11}=
 8\pi P_r B^2
=-\left(\frac{B}{A}\right)^2\left[2\frac{\ddot{R}}{R}-\left(2\frac{\dot
A}{A}-\frac{\dot{R}}{R}\right)
\frac{\dot R}{R}\right]
+\left(2\frac{A^{\prime}}{A}+\frac{R^{\prime}}{R}\right)\frac{R^{\prime}}{R}-\left(\frac{B}{R}\right)^2,
\label{T11} 
\end{eqnarray}
\end{widetext}
\begin{widetext}
\begin{eqnarray}
8\pi T_{22}&=&\frac{8\pi}{\sin^2\theta} T_{33}
=8\pi P_{\perp}R^2
=-\left(\frac{R}{A}\right)^2\left[\frac{\ddot{B}}{B}+\frac{\ddot{R}}{R}
-\frac{\dot{A}}{A}\left(\frac{\dot{B}}{B}+\frac{\dot{R}}{R}\right)
+\frac{\dot{B}}{B}\frac{\dot{R}}{R}\right]
\nonumber \\&+&\left(\frac{R}{B}\right)^2\left[\frac{A^{\prime\prime}}{A}
+\frac{R^{\prime\prime}}{R}-\frac{A^{\prime}}{A}\frac{B^{\prime}}{B}
+\left(\frac{A^{\prime}}{A}-\frac{B^{\prime}}{B}\right)\frac{R^{\prime}}{R}\right],\label{T22}
\end{eqnarray}
\end{widetext}
where dots and primes denote derivative with respect to $t$ and $r$ respectively.
\subsection{Kinematical variables and the mass function}
\label{sec:3}

The three non-vanishing kinematical variables are the four--acceleration  $a_{\alpha}$, the expansion scalar $\Theta$ and the shear tensor $\sigma_{\alpha \beta}$. 
The corresponding expressions follow at once from their definitions. 

Thus
\begin{equation}
a_{\alpha}=V_{\alpha ;\beta}V^{\beta}, \label{4b}
\end{equation}
producing 

\begin{equation}
a_1=\frac{A^{\prime}}{A}, \;\;
a^2=a^{\alpha}a_{\alpha}=\left(\frac{A^{\prime}}{AB}\right)^2,
\label{5c}
\end{equation}
with  $a^\alpha= a \chi^\alpha$.

The expansion $\Theta$ is given by 

\begin{equation}
\Theta={V^{\alpha}}_{;\alpha}=\frac{1}{A}\left(\frac{\dot{B}}{B}+2\frac{\dot{R}}{R}\right),\label{th}
\end{equation}
and for the shear  tensor we have

\begin{equation}
\sigma_{\alpha\beta}=V_{(\alpha
;\beta)}+a_{(\alpha}V_{\beta)}-\frac{1}{3}\Theta h_{\alpha\beta}, \label{4a}
\end{equation}
with only one  non--vanishing independent component

\begin{equation}
\sigma_{11}=\frac{2}{3}B^2\sigma, \;\;
\sigma_{22}=\frac{\sigma_{33}}{\sin^2\theta}=-\frac{1}{3}R^2\sigma,
 \label{5a}
\end{equation}
with
\begin{equation}
\sigma^{\alpha\beta}\sigma_{\alpha\beta}=\frac{2}{3}\sigma^2,
\label{5b}
\end{equation}
being
\begin{equation}
\sigma=\frac{1}{A}\left(\frac{\dot{B}}{B}-\frac{\dot{R}}{R}\right).\label{5b1}
\end{equation}
\subsection{The mass function}
\label{sec:4}
Next, the mass function $m(t,r)$ introduced by Misner and Sharp \cite{14}
is given by 

\begin{equation}
m(t,r)=\frac{R^3}{2}{R_{23}}^{23} 
=\frac{R}{2}\left[\left(\frac{\dot{R}}{A}\right)^2
-\left(\frac{R^{\prime}}{B}\right)^2+1\right].
 \label{18}
\end{equation}

To study the dynamical properties of the system, let us  introduce,
following Misner and Sharp  the proper time derivative $D_T$
given by
\begin{equation}
D_T=\frac{1}{A}\frac{\partial}{\partial t}, \label{16}
\end{equation}
and the proper radial derivative $D_R$,
\begin{equation}
D_R=\frac{1}{R^{\prime}}\frac{\partial}{\partial r}.\label{23a}
\end{equation}

Using (\ref{16}) we can define the velocity $U$ of the collapsing
fluid as the variation of the proper radius with respect to proper time, i.e.\
\begin{equation}
U=D_TR.  \label{19}
\end{equation}
Then (\ref{18}) can be rewritten as
\begin{equation}
E \equiv \frac{R^{\prime}}{B}=\left[1+U^2-\frac{2m(t,r)}{R} \right]^{1/2}.
\label{20}
\end{equation}

From (\ref{18}) we may easily obtain
\begin{eqnarray}
D_Rm=4\pi\left( \mu+  q \frac{U}{E}\right)R^2 .
\label{27b}
\end{eqnarray}
Equation (\ref{27b}) may be integrated to obtain
\begin{equation}
m=\int^{r}_{0}4\pi R^2 \left( \mu +  q \frac{U}{E}\right)R^\prime dr, \label{27int}
\end{equation}
(assuming a regular centre to the distribution, so $m(0)=0$), 
or
\begin{equation}
\frac{3m}{R^3} =4 \pi  \mu - \frac{4 \pi}{R^3}\int^r_0{R^3  \mu^\prime  dr} + \frac{4 \pi}{R^3}\int^r_0{3  q \frac{U}{E}R^2 R^\prime dr} \label{3mi}.
\end{equation}

\subsection{Weyl tensor}
\label{sec:5}
The Weyl tensor is defined through the  Riemann tensor
$R^{\rho}_{\alpha \beta \mu}$, the  Ricci tensor 
$R_{\alpha\beta}$ and the curvature scalar $\cal R$, as:
$$
C^{\rho}_{\alpha \beta \mu}=R^\rho_{\alpha \beta \mu}-\frac{1}{2}
R^\rho_{\beta}g_{\alpha \mu}+\frac{1}{2}R_{\alpha \beta}\delta
^\rho_{\mu}-\frac{1}{2}R_{\alpha \mu}\delta^\rho_\beta$$
\begin{equation}
+\frac{1}{2}R^\rho_\mu g_{\alpha \beta}+\frac{1}{6}{\cal
R}(\delta^\rho_\beta g_{\alpha \mu}-g_{\alpha
\beta}\delta^\rho_\mu). \label{34}
\end{equation}

In the general case the Weyl tensor may be expressed through two tensors denoted as the electric and the magnetic part of the Weyl tensor. In the spherically symmetric case the magnetic part of the Weyl tensor vanishes identically, whereas the electric  part of  Weyl tensor is defined by 
\begin{equation}
E_{\alpha \beta} = C_{\alpha \mu \beta \nu} V^\mu V^\nu,
\label{elec}
\end{equation}
with the following non--vanishing components

\begin{eqnarray}
E_{11}&=&\frac{2}{3}B^2 {\cal E},\nonumber \\
E_{22}&=&-\frac{1}{3} R^2 {\cal E}, \nonumber \\
E_{33}&=& E_{22} \sin^2{\theta}, \label{ecomp}
\end{eqnarray}
where
\begin{widetext}
\begin{eqnarray}
{\cal E}= \frac{1}{2 A^2}\left[\frac{\ddot R}{R} - \frac{\ddot B}{B} - \left(\frac{\dot R}{R} - \frac{\dot B}{B}\right)\left(\frac{\dot A}{A} + \frac{\dot R}{R}\right)\right]
+ \frac{1}{2 B^2} \left[\frac{A^{\prime\prime}}{A} -
\frac{R^{\prime\prime}}{R} + \left(\frac{B^{\prime}}{B} +
\frac{R^{\prime}}{R}\right)\left(\frac{R^{\prime}}{R}-\frac{A^{\prime}}{A}\right)\right]
- \frac{1}{2 R^2}. \label{E}
\end{eqnarray}
\end{widetext}
Observe that we may also write $E_{\alpha\beta}$ as
\begin{equation}
E_{\alpha \beta}={\cal E} \left(\chi_\alpha
\chi_\beta-\frac{1}{3}h_{\alpha \beta}\right). \label{52}
\end{equation}

Using Einstein equations, (\ref{18}) and (\ref{E}) we can write
\begin{equation}
\frac{3m}{R^3}= 4\pi \left( \mu -  P_r + P_ {\perp}\right) - {\cal E}.
\label{mW}
\end{equation}

\subsection{Structure scalars and complexity factor}
\label{sec:6}
The structure scalars are quantities obtained from the orthogonal splitting of the Riemann tensor, which have been shown to play an important role in the study of self--gravitating systems.   They were defined in \cite{15} , and are relevant to our discussion since the variable intended to asses the degree of complexity of  the self--gravitating system (the complexity factor), is one of the structure scalars (see \cite{1}  for details).

Thus, let us define the tensor $Y_{\alpha \beta}$ (the electric part of the Riemann tensor) by
\begin{equation}
Y_{\alpha \beta}=R_{\alpha \gamma \beta \delta}V^\gamma V^\delta.
\label{electric}
\end{equation}

\noindent The tensor $Y_{\alpha \beta}$  may be expressed as 
\begin{eqnarray}
Y_{\alpha\beta}=\frac{1}{3}Y_T h_{\alpha
\beta}+Y_{TF}\left(\chi_{\alpha} \chi_{\beta}-\frac{1}{3}h_{\alpha
\beta}\right).\label{electric'}
\end{eqnarray}
Then from (\ref{T00})--(\ref{T22}) and (\ref{E}) we obtain
\begin{equation}
Y_T=4\pi( \mu+3 P_r-2\Pi), \qquad
Y_{TF}={\cal E}-4\pi \Pi \label{EY},
\end{equation}
and from (\ref{mW}) and (\ref{EY}) we have 
\begin{equation}
Y_{TF}=4 \pi  \mu -8 \pi \Pi - \frac{3m}{R^3},
\label{Ym}
\end{equation}
or, using (\ref{3mi}) we obtain

\begin{equation}
Y_{TF}= - 8 \pi \Pi   + \frac{4 \pi}{R^3}\int^r_0{R^3\left(\mu^\prime - \frac{3  q BU}{R}\right) dr}.
\label{Yi}
\end{equation}

The above equation relates $Y_{TF}$ with the matter variables, however we shall need the expression of this scalar in terms  of  metric and kinematical variables, which  reads (see Eq.(45) in \cite{sins})
\begin{equation}
Y_{TF}\equiv {\cal E} - 4\pi \Pi  = \frac{a^\prime}{B} - \frac{\dot{\sigma}}{A} +a^2 - \frac{\sigma^2}{3} - \frac{2}{3} \Theta \sigma - a \frac{R'}{RB}\;.
\label{shevp}
\end{equation}

As in \cite{1} and \cite{2}, we shall adopt the above scalar  as the variable measuring the degree of complexity of the fluid distribution (the complexity factor). Our models will satisfy the condition of minimal complexity ($Y_{TF}=0$).

\section{THE  JUNCTION CONDITIONS}
\label{sec:2}
If we wish to avoid the  presence of shells on the boundary surfaces delimiting our models, then matching (Darmois) conditions must be imposed. 
Since, as mentioned in the Introduction, some of the obtained models describe fluid distributions with  a void  (cavity) surrounding the center, then in this latter case we have to consider the matching not only on the exterior boundary but on the inner one as well \cite{16}.

Outside $\Sigma^{(e)}$  we have the Vaidya
spacetime (or Schwarzschild in the dissipationless case),
described by
\begin{equation}
ds^2=-\left[1-\frac{2M(v)}{r}\right]dv^2-2drdv+r^2(d\theta^2
+\sin^2\theta
d\phi^2) \label{19d},
\end{equation}
where $M(v)$  denotes the total mass,
and  $v$ is the retarded time.
The matching of the non-adiabatic sphere to
the Vaidya spacetime, on the surface $r=r_{\Sigma^{(e)}}=$ constant, in the absence of thin shells, implies   the continuity of the first and the second fundamental forms through the matching hypersurface, producing
\begin{equation}
m(t,r)\stackrel{\Sigma^{(e)}}{=}M(v), \label{20}
\end{equation}
and
\begin{equation}
q\stackrel{\Sigma^{(e)}}{=}\frac{L}{4\pi r}\stackrel{\Sigma^{(e)}}{=}P_r , \label{20lum}
\end{equation}
where $\stackrel{\Sigma^{(e)}}{=}$ means that both sides of the equation
are evaluated on $\Sigma^{(e)}$ and $L$ denotes   the total luminosity of the  sphere as measured on its surface and is given by
\begin{equation}
L=L_{\infty}\left(1-\frac{2m}{r}+2\frac{dr}{dv}\right)^{-1}, \label{14a}
\end{equation}
where
\begin{equation}
L_{\infty}=\frac{dM}{dv}, \label{14b}
\end{equation}
is the total luminosity measured by an observer at rest at infinity.

In the case when a cavity forms, then we also have to match the solution to the Minkowski spacetime on the boundary surface delimiting the empty cavity ($\Sigma^{(i)}$). In this case the matching conditions imply
\begin{equation}
m(t,r)\stackrel{\Sigma^{(i)}}{=}0, \label{junction1i}
\end{equation}

\begin{equation}
q\stackrel{\Sigma^{(i)}}{=}P_r\stackrel{\Sigma^{(i)}}{=}0.\label{j3in}
\end{equation}

For some models the Darmois conditions cannot be satisfied, in which case we must  allow  the presence of thin shells on $\Sigma^{(i)}$ and/or $\Sigma^{(e)}$, implying   discontinuities in the mass function \cite{17}.

\section{THE QUASI--HOMOLOGOUS CONDITION}
\label{sec:3}
As mentioned before, for time dependent systems it is not enough to define the complexity of the fluid distribution. We need also to elucidate  what is the simplest pattern of evolution of the system.

In \cite{2} it was assumed that the homologous evolution represents the simplest mode of evolution of the fluid distribution. Here we shall relax this condition since it appears to be too stringent thereby  excluding many potential interesting scenarios. Instead we shall assume that the system evolves in a ``quasi--homologous'' regime, whose definition is given below.

First of all let us observe that  we can write  the field equation (\ref{T01}) as
\begin{equation}
D_R\left(\frac{U}{R}\right)=\frac{4 \pi}{E} q+\frac{\sigma}{R},
\label{vel24}
\end{equation}
which can be easily integrated to obtain
\begin{equation}
U=\tilde a(t) R+R\int^r_0\left(\frac{4\pi}{E} q+\frac{\sigma}{R}\right)R^{\prime}dr,
\label{vel25}
\end{equation}
where $\tilde a$ is an integration function, or
\begin{equation}
U=\frac{U_{\Sigma^{(e)}}}{R_{\Sigma^{(e)}}}R-R\int^{r_{\Sigma^{(e)}}}_r\left(\frac{4\pi}{E} q+\frac{\sigma}{ R}\right)R^{\prime}dr.
\label{vel26}
\end{equation}
If the integral in the above equations vanishes  we have from (\ref{vel25}) or (\ref{vel26}) that
\begin{equation}
 U=\tilde a(t) R.
 \label{ven6}
 \end{equation}
 
 This relationship  is characteristic of the homologous evolution in Newtonian hydrodynamics \cite{20,21,22}. In our case, this may occur if the fluid is shear--free and non dissipative, or if the two terms in the integral cancel each other.

In \cite{2}, the term  ``homologous evolution'' was used to characterize  relativistic systems satisfying, besides  (\ref{ven6}), the condition
\begin{equation}
\frac{R_I}{R_{II}}=\mbox{constant},
\label{vena}
\end{equation}
where $R_I$ and $R_{II}$ denote the areal radii of two concentric shells ($I,II$) described by $r=r_I={\rm constant}$, and $r=r_{II}={\rm constant}$, respectively. 

The important point that we want to stress here is that (\ref{ven6}) does not imply (\ref{vena}).
Indeed, (\ref{ven6})  implies that for the two shells of fluids $I,II$ we have 
\begin{equation}
\frac{U_I}{U_{II}}=\frac{A_{II} \dot R_I}{A_I \dot R_{II}}=\frac{R_I}{R_{II}},
\label{ven3}
\end{equation}
that implies (\ref{vena}) only if  $A=A(t)$, which by a simple coordinate transformation becomes $A={\rm constant}$. Thus in the non--relativistic regime, (\ref{vena}) always follows from the condition that  the radial velocity is proportional to the radial distance, whereas in the relativistic regime the condition (\ref{ven6}) implies (\ref{vena}), only if the fluid is geodesic. 

We shall define  quasi--homologous evolution as that restricted only by condition  (\ref{ven6}), implying 
\begin{equation}
\frac{4\pi}{R^\prime}B  q+\frac{\sigma}{ R}=0.
\label{ch1}
\end{equation}

Thus our models will be restricted by (\ref{ch1})  and  $Y_{TF}=0$.
\section{THE TRANSPORT EQUATION}
\label{sec:4}
 In the diffusion approximation we shall need a transport equation to evaluate the temperature and its evolution within the fluid distribution. Here we shall resort to a transport equation derived from a causal  dissipative theory ( e.g. the
M\"{u}ller-Israel-Stewart second
order phenomenological theory for dissipative fluids \cite{Muller67,IsSt76,I,II}).

Indeed, as it is already  well known the Maxwell-Fourier law for
heat flux leads to a parabolic equation (diffusion equation)
which predicts propagation of perturbations with infinite speed
(see \cite{6D}-\cite{8'} and references therein). This simple fact
is at the origin of the pathologies \cite{9H} found in the
approaches of Eckart \cite{10E} and Landau \cite{11L} for
relativistic dissipative processes. To overcome such difficulties,
various relativistic
theories with non-vanishing relaxation times have been proposed in
the past \cite{Muller67,IsSt76,I,II,14Di,15d}. Although the final word on this issue has not yet been said, the important point is that
all these theories provide a heat transport equation which is not
of Maxwell-Fourier type but of Cattaneo type \cite{18D}, leading
thereby to a hyperbolic equation for the propagation of thermal
perturbations.

In all these theories the relaxation time $\tau$ is not neglected, allowing them to study  transient regimes.

It is worth mentioning that  large relaxation times (large
mean free paths of particles involved in heat transport) does not imply 
departure from the hydrodynamic regime, since the latter is related to the mean free path of the particles forming the fluid, which in general are different from those  responsible for the heat transport, (this fact has been streseed before
\cite{Santos}, but it is usually overlooked).

Thus the  corresponding  transport equation for the heat flux reads
\begin{equation}
\tau
h^{\alpha\beta}V^{\gamma}q_{\beta;\gamma}+q^{\alpha}=-\kappa h^{\alpha\beta}
(T_{,\beta}+Ta_{\beta}) -\frac 12\kappa T^2\left( \frac{\tau
V^\beta }{\kappa T^2}\right) _{;\beta }q^\alpha ,  \label{21t}
\end{equation}
where $\kappa $  denotes the thermal conductivity, and  $T$ and
$\tau$ denote temperature and relaxation time respectively. Observe
that, due to the symmetry of the problem, equation (\ref{21t}) only
has one independent component, which may be written as
\begin{equation}
\tau{\dot q}=-\frac{1}{2}\kappa qT^2\left(\frac{\tau}{\kappa
T^2}\right)^{\dot{}}-\frac{1}{2}\tau q\Theta A-\frac{\kappa}{B}(TA)^{\prime}-qA.
\label{te}
\end{equation}
In the case $\tau=0$ we recover the Eckart--Landau equation.

 For simplicity we shall consider here the so called ``truncated'' version where the last term in (\ref{21t}) is neglected \cite{19n},
\begin{equation}
\tau
h^{\alpha\beta}V^{\gamma}q_{\beta;\gamma}+q^{\alpha}=-\kappa h^{\alpha\beta}
(T_{,\beta}+Ta_{\beta}) \label{V1},
\end{equation}
and whose    only non--vanishing independent component becomes
\begin{equation}
\tau \dot q+qA=-\frac{\kappa}{B}(TA)^{\prime}. \label{V2}
\end{equation}

\section{ANOTHER  DEFINITION OF RADIAL VELOCITY AND SOME KINEMATICAL RESTRICTIONS}
\label{sec:6}
In order to obtain our models, besides the condition of the vanishing complexity factor and the quasi--homologous evolution, we need to impose further conditions on the system. Here we analyze additional restrictions  on some kinematical variables. For doing that let us first   introduce another concept of velocity, different from $U$.

In the previous section we defined  the  variable $U$ which, as mentioned before, measures the variation of the areal radius $R$  per unit of proper time.  However, another possible definition of ``velocity'' may be introduced, as    the variation of the infinitesimal proper radial distance between two neighboring points ($\delta l$) per unit of proper time, i.e. $D_T(\delta l)$.
Thus, it can be shown that (see \cite{16,18,19} for details)
\begin{equation}
 \frac{D_T(\delta l)}{\delta l}= \frac{1}{3}(2\sigma +\Theta),
\label{vel15}
\end{equation}
or,
\begin{equation}
 \frac{D_T(\delta l)}{\delta l}= \frac{\dot B}{AB}.
\label{vel16}
\end{equation}
Then  we can write
\begin{equation}
 \sigma= \frac{D_T(\delta l)}{\delta l}-\frac{D_T R}{R}= \frac{D_T(\delta l)}{\delta l}-\frac{U}{R},
\label{vel17}
\end{equation}
and
\begin{equation}
 \Theta= \frac{D_T(\delta l)}{\delta l}+\frac{2D_T R}{R}=\frac{D_T(\delta l)}{\delta l}+\frac{2U}{R},
\label{vel17bis}
\end{equation}
The  ``areal'' velocity $U$,  is  related to the change  of areal radius $R$  of a layer of matter, whereas $D_T(\delta l)$, has also the meaning of ``velocity'', being the relative velocity between neighboring layers of matter, and is  in general different from $U$.

In  \cite{19} it was shown that the condition $\Theta=0$ requires the existence of a cavity surrounding the centre of the fluid distribution. There are however another kinematical conditions compatible with the formation of a cavity around the center of symmetry \cite{16}.

Indeed, let us consider the condition  $D_T(\delta l)=0$, but $U\neq 0$. From the  comments above it is evident why we shall refer to it as  the purely areal evolution condition.

Now,   if $D_T(\delta l)=0$ then  $B=B(r)$ from which  a reparametrization of the coordinate $r$ allows us to write without loss of generality $B=1$ implying $R^\prime=E$, and as it follows from (\ref{vel17}) and (\ref{vel17bis})
\begin{equation}
\sigma=-\frac{U}{R}=-\frac{\Theta}{2}
\label{nk1}
\end{equation}
Then, we can write   (\ref{vel24}) in the form

\begin{equation}
\sigma^{\prime}+\frac{\sigma R^{\prime}}{R}=-4\pi q,
\label{27}
\end{equation}
whose integration with respect to $r$ yields
\begin{equation}
\sigma=\frac{\zeta(t)}{R}-\frac{4\pi}{R}\int^{r}_{0} qRdr,
\label{28relvel}
\end{equation}
where $\zeta$ is an integration function of  $t$. It should be observed that in the case where the fluid fills all the sphere, including the centre ($r=0$), we should impose the regularity condition $\zeta=0$. However since we consider the possibility of  a cavity surrounding the centre, such a condition is not required.

From  (\ref{28relvel}) it follows that
\begin{equation}
U=-\zeta+4\pi\int^{r}_{0} qRdr.
\label{29relvel}
\end{equation}
Let us notice that the expression above is compatible with (\ref{ven6}) and (\ref{ch1}). Indeed, taking the  $r$ derivative of (\ref{29relvel}) and using (\ref{ch1}) we obtain (\ref{ven6}). Or, taking the $r$ derivative of (\ref{29relvel})  and using (\ref{ven6}) we obtain (\ref{ch1}).

 Then assuming   the purely areal evolution condition,  if the fluid fills the whole sphere (no cavity surrounding the centre),  and we have a symmetry centre,  we have to put $\zeta=0$, and (\ref{29relvel}) becomes
\begin{equation}
U=4\pi\int^{r}_{0} qRdr.
\label{29relvelbis}
\end{equation}

On the other hand,  if the centre is surrounded by  a compact spherical section of another spacetime, suitably matched to the rest of the fluid, e.g.  if we  choose an inner vacuum Minkowski spherical vacuole then $\zeta$ may be different from zero. This latter case will be considered here for reasons that we explain below.

The point is that   the appearance of a cavity under the assumed conditions is suggested by  (\ref{29relvelbis}). Indeed, in the case of an outwardly directed flux vector ($q>0$), all terms within the integral are positive and  we obtain from (\ref{vel17bis}) and (\ref{29relvelbis}) that $\Theta>0$ and $U>0$. Now, during the Kelvin-Helmholtz phase of evolution  \cite{21}, when all the dissipated energy comes from the gravitational energy, we should expect  a  contraction, not an expansion, to be associated with an outgoing dissipative flux. Inversely,  an inwardly directed flux ($q<0$)  (during that phase) would produce an overall expansion instead of a contraction as it follows from (\ref{29relvelbis}). The above comments suggest  that $\zeta\neq0$.

Thus we have seen that the purely areal evolution condition appears to be  particularly suitable to describe  the evolution of a fluid distribution with  a cavity surrounding the centre.

Another possible restriction on the kinematics of the fluid is provided by the case $U=0$ but $D_T(\delta l)\neq0$. Thus the areal radius remains constant but  the infinitesimal proper radial distance between two neighboring points changes with time. 
As strange as this case might look like, we were unable to   rule it out by mathematical or physical arguments, and therefore we shall consider solutions satisfying such a condition.
Then it follows at once from (\ref{vel16}) and (\ref{vel17}) that, for this latter case,  the condition of quasi--homologous evolution (\ref{ch1}) becomes
\begin{equation}
\frac{\dot B}{AB}=-\frac{4\pi Rq}{E},\qquad \Theta=\sigma.
\label{vu1}
\end{equation}
It is worth noticing that this kinematical condition does not force the formation of a cavity  surrounding the center.

In the next section we shall consider some models satisfying conditions $Y_{TF}=0$, (\ref{ch1}) and either (\ref{29relvel}) or (\ref{vu1}).

\section{MODELS}
\label{sec:7}

In what follows we shall present some exact solutions describing  either  dissipative or non--dissipative systems satisfying the vanishing complexity factor condition and evolving quasi--homologously. Further conditions shall be necessary in order to fully specify the models, these additional restrictions will be based on the material exposed in the previous section.

\subsection{Non dissipative models.}
\label{sec:8}

Although in this work we are mainly concerned  with dissipative systems, for the sake of completeness  we shall first consider the non--dissipative case $q=0$. 

In  \cite{2}  it was shown that  in this case  the homologous condition  (\ref{ven6}) and (\ref{vena}) implies $Y_{TF}=0$ and the fluid is geodesic. Furthermore, there is a unique model evolving homologously and satisfying $Y_{T F} = 0$,  (Friedman--Robertson--Walker). 

We shall  now explore the situation when  the system evolves under the quasi--homologous condition (condition (\ref{ven6}) is satisfied but  (\ref{vena})  is not).

From (\ref{ch1}) it follows at once that $q=0$ implies  $\sigma=0$, i.e. the fluid is shear--free. This last condition implies that 
\begin{equation}
R=rB,
\label{nd1}
\end{equation}
using the above equation we can write (\ref{T01}) as 
\begin{equation}
\left(\frac{\dot B}{AB}\right)^\prime=0.
\label{nd2}
\end{equation}

On the other hand the quasi--homologous condition (\ref{ven6}) may be written as
\begin{equation}
U=\frac{\dot R}{A}=\frac{r\dot B}{A}=\tilde a(t)r B,
\label{nd3}
\end{equation}
which obviously satisfies (\ref{nd2}). In other words condition (\ref{nd3}) is compatible with the field equations.

It should be noticed that in this case either restriction $B=1$, or  $U=0$ would produce a static model. Also, the condition that the fluid is geodesic would lead to the Friedman--Robertson--Walker as  in \cite{2}.

 Next we have to impose the vanishing  complexity factor condition $Y_{TF}=0$, which using (\ref{shevp}) and the condition $\sigma=0$ reads
 \begin{equation}
 \frac{a^\prime}{B}+a^2-\frac{aR^\prime}{BR}=0,
\label{q04}
\end{equation}
or, using (\ref{5c})
\begin{equation}
A^{\prime \prime }-\frac{2A^\prime B^\prime}{B}-\frac{A^\prime}{r}=0,
\label{ndisn1}
\end{equation}
which may be integrated producing
\begin{equation}
A^\prime=BRF(t)=\frac{R^2}{r}F(t),
\label{q04n}
\end{equation}
where $F(t)$ is an arbitrary function of integration.

Next, using (\ref{nd1}), (\ref{nd3}) and (\ref{q04n}), we may rewrite (\ref{nd2}) as 
\begin{equation}
\frac{\dot R^\prime}{R}-\frac{\dot R R^\prime}{R^2}=\frac{R^2}{r}\tilde a(t)F(t).
\label{q07}
\end{equation}

Introducing  $y=\frac{R^\prime}{R}$, (\ref{q07}) can be written as
\begin{equation}
\dot y^\prime=\dot y \left(2y-\frac{1}{r}\right),
\label{q071}
\end{equation}
which may be transformed further by defining the intermediate  variable  $z=yr$,  producing
\begin{equation}
\dot z^\prime=\frac{2z\dot z}{r},
\label{q072}
\end{equation}
or, introducing the independent variable $x=\ln r$, we can finally write eq.(\ref{q072}) as 
\begin{equation}
\frac{d\dot z}{dx}=2z\dot z.
\label{q073}
\end{equation}
Using Mathematica to integrate this last equation  we obtain
\begin{equation}
z=-c_2\tanh(c_1 t + c_2 \ln r + c_3),
\label{q0z}
\end{equation}
and using this expression we have 
\begin{equation}
R=\frac{\tilde R(t)}{\cosh(c_1 t + c_2 \ln r + c_3)},
\label{q0R}
\end{equation}
where $\tilde R(t)$ is an arbitrary function of integration and $c_1, c_2, c_3$ are  arbitrary constants of integration.

To specify further our model we shall assume  $\tilde a(t)=\tilde a=$constant, in which case the physical variables become

\begin{equation}
8\pi\mu=3\tilde a^2+\frac{1}{\tilde R^2(t)}\left[3c_2^2+(1-c_2^2)\cosh^2u\right],
\label{q0mu}
\end{equation}
\begin{widetext}
\begin{eqnarray}
8\pi P_r=-3\tilde a^2+\frac{1}{\tilde R^2(t)D}\left\{\dot{\tilde R}(t)\left[-c_2^2+(c_2^2-1)\cosh^2u\right] 
+ \tilde R(t) c_1 \tanh u\left[3c_2^2+(1-c_2^2)\cosh^2u\right]\right\},
\label{q0pr}
\end{eqnarray}
\end{widetext}
\begin{equation}
8\pi P_\bot=-3\tilde a^2+\frac{c_2^2}{\tilde R^2(t)D}\left(3\tilde R(t)c_1 \tanh u-\dot{\tilde R}(t)\right),
\label{q0pt}
\end{equation}
where 
\begin{equation}
u=c_1 t + c_2 \ln r + c_3,
\label{q0u}
\end{equation}
and
\begin{equation}
D\equiv \dot{\tilde R}(t)-\tilde R(t) c_1 \tanh u.
\label{q0E}
\end{equation}

It is a simple matter to check that for a wide range of values of the parameters, the above solution is singular--free, and satisfies the usual energy  conditions as well as the boundary conditions  (e.g. $c^2_2=1, c_1<0$). However  we are not interested in a particular model, but just want to illustrate the point that once the homologous condition is relaxed and one assumes the quasi--homologous  one, then a great deal of models satisfying the condition $Y_{TF} =0$ are available.
\subsection{Dissipative models with $D_T(\delta l)=0$, $U\neq0$.}
\label{sec:9}
\noindent We shall now consider models satisfying the constraint $D_T(\delta l)=0$, which as mentioned before implies $B=1$. These models are endowed with a cavity surrounding the center, accordingly we should not worry about regularity  conditions at the centre.

In this case the physical variables read

\begin{equation}
  8\pi \mu = \frac{1}{A^2}\frac{\dot{R}^2}{R^2}-\frac{2R^{\prime\prime}}{R}-\frac{R^{\prime 2}}{R^2}+\frac{1}{R^2}, \label{mu}
  \end{equation}

  \begin{equation}\label{Pr}
    8\pi P_r = -\frac{1}{A^2}\left( \frac{2\ddot{R}}{R}-\frac{2\dot{A}}{A}\frac{\dot{R}}{R}+\frac{\dot{R}^2}{R^2}\right)
  +\frac{2A^\prime}{A}\frac{R^\prime}{R}+\frac{{R^{\prime}}^2}{R^2}-\frac{1}{R^2},
  \end{equation}

  \begin{equation}\label{Pt}
   8\pi P_\perp =  -\frac{1}{A^2}\left ( \frac{\ddot{R}}{R}-\frac{\dot{A}}{A}\frac{\dot{R}}{R}\right )+\frac{A^{\prime\prime}}{A}+\frac{R^{\prime\prime}}{R}+\frac{A^\prime}{A}\frac{R^\prime}{R},
  \end{equation}
  \begin{equation}\label{fq}
  4\pi q =\frac{1}{A}\left (\frac{\dot{R}^\prime}{R}-\frac{A^\prime}{A}\frac{\dot{R}}{R}\right ) = -\sigma \frac{R^\prime}{R},
\end{equation}
and for the kinematical variables we have

\begin{equation}\label{VaCi}
\sigma =-\frac{\dot{R}}{AR},\qquad \Theta=\frac{2\dot{R}}{AR}.
\end{equation}

Next, imposing the quasi--homologous condition, we obtain

\begin{equation}\label{CoHo}
  U=\tilde{a}(t) R=\frac{\dot{R}}{A}\qquad \Rightarrow\qquad \tilde{a}(t)=\frac{\dot{R}}{AR}\qquad\Rightarrow \qquad \sigma=-\tilde{a}(t),
\end{equation}
\begin{equation}\label{DCoHo}
   \Theta-\sigma=3\tilde{a}(t).
\end{equation}

On the other hand the condition $Y_{TF}=0$ produces 
\begin{eqnarray}
  Y_{TF} &=& \frac{1}{A^2} \left(\frac{\ddot{R}}{R}-\frac{\dot{A}}{A}\frac{\dot{R}}{R}\right)+\frac{A^{\prime \prime}}{A}-\frac{A^\prime}{A}\frac{R^\prime}{R}\\
   &=&\sigma^2-\frac{\dot{\sigma}}{A} +\frac{A^{\prime \prime}}{A}-\frac{A^\prime}{A}\frac{R^\prime}{R}=0.
\end{eqnarray}

\noindent Thus for this particular case, the conditions of vanishing complexity factor and quasi--homologous evolution read
\begin{equation}\label{Cond1}
  A^{\prime\prime}-\frac{A^\prime R^\prime}{R}+A\sigma^2=\dot{\sigma},
\end{equation}
and
\begin{equation}\label{Cond2}
  \frac{\dot{R}}{R}=-\sigma A,
\end{equation}
respectively.

It would be useful to introduce the intermediate variables $(X, Y)$,
\begin{equation}\label{ccam}
  A=X+\frac{\dot{\sigma}}{\sigma^2}\quad  {\rm and} \quad R=X^\prime Y,
\end{equation}

\noindent in terms of which  (\ref{Cond1}) and  (\ref{Cond2}), become

\begin{equation}\label{Cond1P}
  -\frac{X^\prime}{X}\frac{Y^\prime}{Y}+\sigma^2=0,
\end{equation}

\begin{equation}\label{Cond2P}
  \frac{\dot X^\prime}{X^\prime}+\frac{\dot Y}{Y}=-\sigma X-\frac{\dot{\sigma}}{\sigma}.
\end{equation}

In what follows we shall analyze different models satisfying the above conditions, by imposing additional restrictions.
\subsubsection{Subcase with $X=\tilde{X}(r) {\cal T}(t)$}

\noindent In this first subcase we assume the function $X$ to be separable, i.e.
\begin{equation}\label{VSX}
  X=\tilde X(r) {\cal T}(t).
\end{equation}
\noindent Then  feeding back (\ref{VSX}) into (\ref{Cond1P}), and taking $t$-derivative we obtain
\begin{equation}\label{Cond1PV}
  -\frac{\tilde{X^\prime}}{\tilde{X}}\left (\frac{\dot Y}{Y}\right )^\prime+2\sigma\dot{\sigma}=0.
\end{equation}

\noindent Likewise, feeding back (\ref{VSX}) into  (\ref{Cond2P}) and taking the $r$-derivative we obtain
\begin{equation}\label{Cond2PV}
  \left (\frac{\dot Y}{Y}\right )^\prime=-\sigma \tilde{X}^\prime {\cal T}.
\end{equation}

\noindent The combination of (\ref{Cond1PV}) and (\ref{Cond2PV}) produces
\begin{equation}\label{Cond12PV}
  \frac{\tilde X^{\prime 2}}{\tilde X}=-\frac{2\dot{\sigma}}{{\cal T}}\equiv \beta^2,
\end{equation}
\noindent  where $\beta$ is a constant. 

Then, from the integration of (\ref{Cond12PV}) we have
\begin{equation}\label{solPV}
  \tilde{X}=\frac{(\beta r+c_1)^2}{4}\quad {\rm and}  \quad {\cal T}(t)=-\frac{2\dot{\sigma}}{\beta ^2},
\end{equation}
where $c_1$ is a constant of integration.

\noindent  Thus, the metric functions for this subcase become
\begin{eqnarray}
  A &=& \frac{\dot{\sigma}}{2\beta^2\sigma^2} \left [2\beta^2-\sigma^2 (\beta r+c_1)^2\right],\label{fume1}\\
  R &=& \tilde{R}(t)\frac{\beta}{2}(\beta r+c_1)e^{\frac{\sigma ^2 r}{4\beta}(\beta r+2c_1)},\label{fume}
\end{eqnarray}
\noindent where $\tilde{R}(t)$ is an arbitrary function of time.

\noindent Finally  using  the expressions above in (\ref{mu})-(\ref{fq}) we find  for the physical variables
\begin{widetext}
\begin{equation}\label{muvs}
  8\pi \mu=-3\sigma^2-\frac{3\sigma^4}{4\beta^2}(\beta r+c_1)^2-\frac{\beta^2}{(\beta r+c_1)^2}+\frac{\sigma ^2}{(\beta r+c_1)^2}e^{-\frac{\sigma ^2}{2\beta^2}(\beta r+c_1)^2},
\end{equation}

\begin{equation}\label{Prvs}
  8\pi P_r=-\frac{4\sigma^2\beta^2}{2\beta^2-\sigma^2(\beta r+c_1)^2}+\frac{\beta^2}{(\beta r+c_1)^2}+\frac{\sigma^4}{4\beta^2}(\beta r+c_1)^2-\frac{\sigma ^2}{(\beta r+c_1)^2}e^{-\frac{\sigma ^2}{2\beta^2}(\beta r+c_1)^2},
\end{equation}

\begin{equation}\label{Ptvs}
  8\pi P_\bot=\frac{\sigma^2}{2}+\frac{\sigma^4(\beta r+c_1)^2}{4\beta^2}-\frac{\sigma^2[2\beta^2+\sigma^2(\beta r+c_1)^2]}{2\beta^2-\sigma^2(\beta r+c_1)^2},
\end{equation}
\end{widetext}
\begin{equation}\label{fqvs}
  4\pi q=-\frac{\sigma[2\beta^2+\sigma^2(\beta r+c_1)^2]}{2\beta(\beta r+c_1)}.
\end{equation}

\noindent  Let us now consider the possible matching of this model on $\Sigma^{(i)}$ and $\Sigma^{(e)}$. On the former surface, we must have  $q=0$, implying   $\sigma=0$, on that surface, however since $\sigma $ is only function on $t$, this implies $\sigma=0$ producing a non--dissipative solution.  On  the other hand,  from the matching condition $q\stackrel{\Sigma^{(e)}}{=}P_r$, (\ref{Prvs}) and  (\ref{fqvs}) we obtain
\begin{equation}
\frac{\sigma(2+\sigma^2 K^2)}{K}-\frac{4\sigma^2}{2-\sigma^2 K^2}+\frac{1}{K^2}+\frac{\sigma^4 K^2}{4}-\frac{\sigma^2}{\beta^2 K^2}e^{-\frac{\sigma^2 K^2}{2}}=0,\label{sigmae}
\end{equation}
\noindent where $K=\frac{\beta r_{\Sigma^{(e)}}+c_1}{\beta}$. 

The algebraic equation above only allows solutions for  constant values of $\sigma$ depending on $K,\beta$, i.e. only for fixed values of $t$.  Therefore this model has thin shells on either boundary surfaces $\Sigma^{(i)}$ and $\Sigma^{(e)}$.

Finally, we may calculate the temperature for this model. Using (\ref{V2}), (\ref{fume1})  and (\ref{fqvs}) we obtain
\begin{widetext}
\begin{equation}\label{Temvs}
  T(t,r)=\frac{\beta ^2 \sigma^2}{2\pi \kappa [2\beta^2-\sigma^2(\beta r+c_1)^2]}\left [\left(\tau+ \frac{1}{ \sigma}\right)\ln(\beta r +c_1)+\frac{3\tau \sigma^2}{4\beta^2}(\beta r +c_1)^2-\frac{\sigma^3}{16\beta^4}(\beta r +c_1)^4\right ]+T_0(t),
\end{equation}
\end{widetext}
where $T_0(t)$ is an arbitrary function related to  the temperature at either one of the boundary surfaces.

\subsubsection{Subcase with $A=A(r)$.}

\noindent  In this subcase we assume that 
\begin{equation}\label{Ar}
  A=A(r).
\end{equation}

\noindent Then, taking the $t$-derivative of (\ref{Cond1}) we have
\begin{equation}\label{Cond1Ar}
  -A^\prime \left(\frac{\dot{R}}{R}\right )^\prime+2A\sigma \dot{\sigma}=\ddot{\sigma},
\end{equation}

\noindent whereas the $r$- derivative of (\ref{Cond2}) produces
\begin{equation}\label{Cond2Ar}
   \left(\frac{\dot{R}}{R}\right )^\prime=-\sigma A^\prime.
\end{equation}

\noindent Combining (\ref{Cond1Ar}) and  (\ref{Cond2Ar}) we obtain
\begin{equation}\label{Cond12Ar}
  \sigma (A^\prime)^2+2A\sigma \dot{\sigma}=\ddot{\sigma},
\end{equation}
\noindent whose solution satisfying (\ref{CoHo}) and  (\ref{Ar}) is
\begin{equation}\label{SiAr}
  \sigma =-\sigma_0 t+\sigma_1\qquad (\sigma_0,\,\,\sigma_1\,\, {\rm constants}),
\end{equation}
\noindent and
\begin{equation}\label{AAr}
  \frac{(A^\prime)^2}{A}=2\sigma_0,\qquad A=\frac{1}{4}(\sqrt{2\sigma_0}r +c_1)^2.
\end{equation}

\noindent From the above we find the expression for $R$ which reads
\begin{equation}\label{RAr}
  R=\tilde{R}(r)e^{-\frac{1}{4}(\sqrt{2\sigma_0}r +c_1)^2 (-\frac{\sigma_0}{2}t^2+\sigma_1 t)},
\end{equation}
where $\tilde R(r)$ is an arbitrary function of its argument.

\noindent To obtain a specific model, we shall assume $\tilde R={\rm constant}$, then feeding back  (\ref{AAr}) and  (\ref{RAr})  into  (\ref{mu})-(\ref{fq}) we find for the physical variables

\begin{eqnarray}\label{muAr}
  8\pi \mu&=&\sigma^2_1-\frac{3\sigma_0}{2}(\sqrt{2\sigma_0}r +c_1)^2 \left(-\frac{\sigma_0}{2}t^2+\sigma_1 t\right)^2\nonumber \\&+&\frac{1}{\tilde{R}^2}e^{\frac{1}{2}(\sqrt{2\sigma_0}r +c_1)^2 (-\frac{\sigma_0}{2}t^2+\sigma_1 t)},
\end{eqnarray}

\begin{eqnarray}\label{PrAr}
  8\pi P_r&=&-3\sigma^2_1+2\sigma_0\sigma_1 t-\sigma_0^2 t^2-\frac{8\sigma_0}{(\sqrt{2\sigma_0}r+c_1)^2}+      \frac{\sigma_0}{2}(\sqrt{2\sigma_0}r +c_1)^2 \left(-\frac{\sigma_0}{2}t^2+\sigma_1 t\right)^2\nonumber \\
  &-&\frac{1}{\tilde{R}^2}e^{\frac{1}{2}(\sqrt{2\sigma_0}r +c_1)^2 (-\frac{\sigma_0}{2}t^2+\sigma_1 t)},
\end{eqnarray}

\begin{equation}\label{PtAr}
  8\pi P_\bot=\frac{\sigma_0^2}{2}t^2-\sigma^2_1-\sigma_0\sigma_1 t+\frac{\sigma_0}{2}(\sqrt{2\sigma_0}r +c_1)^2 \left(-\frac{\sigma_0}{2}t^2+\sigma_1 t\right)^2,
\end{equation}

\begin{equation}\label{QAr}
  4\pi q=\frac{\sqrt{2\sigma_0}}{2} (\sqrt{2\sigma_0}r +c_1) \left(-\frac{\sigma_0}{2}t^2+\sigma_1 t\right)(-\sigma_0 t+\sigma_1).
\end{equation}

As in previous models it is a simple matter to check that a  wide range of values of the  parameters allows to construct  singular--free models satisfying the usual energy conditions.

Let us now check the possibility to satisfy the Darmois conditions on $\Sigma^{(i)}$ and/or  $\Sigma^{(e)}$. 

Since regularity  conditions on the radial pressure (\ref{PrAr}) require $(\sqrt{2\sigma_0}r+c_1)\neq 0$, then the matching condition $q\stackrel{\Sigma^{(i)}}{=}P_r\stackrel{\Sigma^{(i)}}{=}0$, with  (\ref{QAr}),  implies  $\sigma_0=0$, producing a non--dissipative solution.

For the exterior boundary surface $\Sigma^{(e)}$,  it is imposible to match to the exterior metric, for  any possible $\tilde R=\tilde R(r)$. 

Thus in this model, both the interior and the exterior boundary surfaces present a thin shell.

Finally, for the temperature of this model we obtain,  using  (\ref{V2}), (\ref{AAr}) and (\ref{QAr})
\begin{widetext}
\begin{equation}
  T(t,r) = -\frac{1}{4\pi \kappa}\left [ \tau\left(\frac{3}{2}\sigma_0^2 t^2-3\sigma_0\sigma_1 t+\sigma_1^2\right)
   +\frac{(\sqrt{2\sigma_0}r+c_1)^2}{8}\left(-\frac{\sigma_0}{2}t^2+\sigma_1t\right)(-\sigma_0 t+\sigma_1) \right ]+T_0(t).\label{TemAr}
\end{equation}
\end{widetext}
\subsubsection{Subcase with  $\dot{\sigma}=0$}

\noindent We shall here present a model satisfying the additional condition
\begin{equation}\label{sigcons}
  \dot{\sigma}=0\Rightarrow \sigma={\rm constant}.
\end{equation}

\noindent Next, introducing the intermediate variable $Y$ defined by
\begin{equation}\label{CamSig}
  R=A^\prime Y,
\end{equation}
\noindent we may write  (\ref{Cond1}) and  (\ref{Cond2}) as
\begin{equation}\label{Cond1Sig}
  \frac{A^\prime}{A}\frac{Y^\prime}{Y}=\sigma^2,
\end{equation}
\noindent and 
\begin{equation}\label{Cond2Sig}
  \frac{\dot{A}^\prime}{A^\prime}+\frac{\dot{Y}}{Y}=-\sigma A.
\end{equation}

\noindent Then taking the $t$-derivative of (\ref{Cond1Sig})  and the $r$-derivative of (\ref{Cond2Sig}) we obtain
\begin{equation}\label{Cond1Sig1}
  \left (\frac{\dot{Y}}{Y}\right )^\prime=-\left (\frac{\dot{A}}{A}\right )^\prime \sigma^2\left (\frac{A}{A^\prime}\right )^2,
\end{equation}

\noindent and 
\begin{equation}\label{Cond2Sig2}
  \left (\frac{\dot{Y}}{Y}\right )^\prime=-\sigma A^\prime-\left (\frac{\dot{A}^\prime}{A^\prime}\right )^\prime,
\end{equation}
the combination of which produces
\begin{equation}\label{Cond12Sig}
  \sigma^2(A^\prime \dot A-\dot A^\prime A )+\sigma A^{\prime 3}+\dot A^{\prime\prime}A^\prime -\dot{A}^\prime A^{\prime\prime}=0.
\end{equation}

\noindent A solution to the  equation (\ref{Cond12Sig}) is
\begin{equation}\label{SolSig}
  A=\beta r-\frac{\beta^2}{\sigma}t+\beta_0,
\end{equation}
\noindent  producing for $R$ 
\begin{equation}\label{RSig}
  R=\tilde{R}_0\beta e^{(\frac{\sigma^2}{2}r^2-\sigma \beta t r+\frac{\sigma^2 \beta_0}{\beta} r+\frac{\beta^2}{2}t^2-\sigma\beta_0 t)}\quad (\tilde{R_0}, \beta, \beta_0={\rm const}).
\end{equation}

\noindent Using (\ref{SolSig}) and  (\ref{RSig}) in  (\ref{mu})-(\ref{fq}) the physical variables read for this subcase
\begin{eqnarray}\label{muSig}
   8\pi \mu= -\sigma ^2-3\left [ \sigma^2 \left(r+\frac{\beta_0}{\beta}\right)-\sigma \beta t\right ]^2 \nonumber \\+\frac{e^{-2(\frac{\sigma^2}{2}r^2-\sigma \beta t r+\frac{\sigma^2 \beta_0}{\beta} r+\frac{\beta^2}{2}t^2-\sigma\beta_0 t)}}{(\tilde{R}_0\beta )^2},
\end{eqnarray}

\begin{eqnarray}\label{PrSig}
  8\pi P_r=-\sigma^2+\sigma^4\left (  r-\frac{\beta}{\sigma}t +\frac{\beta_0}{\beta}\right)^2\nonumber \\-\frac{e^{-2(\frac{\sigma^2}{2}r^2-\sigma \beta t r+\frac{\sigma^2 \beta_0}{\beta} r+\frac{\beta^2}{2}t^2-\sigma\beta_0 t)}}{(\tilde{R}_0\beta )^2},
\end{eqnarray}

\begin{equation}\label{PtSig}
  8\pi P_\bot=\sigma^2+\left [\sigma^2\left(r+\frac{\beta_0}{\beta}\right)-\sigma\beta t \right]^2,
\end{equation}

\begin{equation}\label{QSig}
  4\pi q=-\sigma^3 \left (r-\frac{\beta}{\sigma}t+\frac{\beta_0}{\beta} \right ).
\end{equation}

In this case is a simple matter to check that Darmois conditions cannot be satisfied on either boundary surface $\Sigma^{(i)}$ or  $\Sigma^{(e)}$.

For the temperature of this model we obtain
\begin{equation}
  T(t,r) = -\frac{\beta \sigma^2}{4\pi \kappa(\beta r-\frac{\beta^2}{\sigma}t+\beta_0)}\left [ \tau r-\frac{\sigma}{3}\left(r-\frac{\beta}{\sigma}t+\frac{\beta_0}{\beta}\right)^3 \right ]+T_0(t).\label{TemSig}
\end{equation}
We shall next, turn to our last family of models, characterized by a vanishing $U$.

\subsection{Dissipative models with $D_T(\delta l)\neq0$ and $U=0$}
Assuming $U=0$, it follows that 
\begin{equation}
U=\frac{\dot R}{A}=0 \Rightarrow R=R(r).
\label{U0}
\end{equation}
Then the following condition  applies to the kinematical variables
\begin{equation}
\sigma=\Theta=\frac{\dot B}{AB}.
\label{U0sigthe}
\end{equation}

Using the conditions above, together with the  quasi--homologous condition (\ref{ch1}),  the vanishing of the complexity factor $Y_{TF}=0$  (eq.  (\ref{shevp})), becomes
\begin{equation}
\frac{1}{A^2}\left(\frac{\ddot B}{B}-\frac{\dot A \dot B}{AB}\right)=\frac{1}{B^2}\left(\frac{A^{\prime\prime}}{A}-\frac{A^\prime B^\prime}{AB}-\frac{A^\prime R^\prime}{AR}\right).
\label{YTF0}
\end{equation}

We shall now construct a model satisfying the above equation, by assuming

\begin{equation}
\frac{\ddot B}{B}-\frac{\dot A \dot B}{AB}=0,
\label{YTF0i}
\end{equation}
\begin{equation}
\frac{A^{\prime\prime}}{A}-\frac{A^\prime B^\prime}{AB}-\frac{A^\prime R^\prime}{AR}=0,
\label{YTF0d}
\end{equation}
the integration of which produces
\begin{equation}
\dot B=A f(r),
\label{CI1}
\end{equation}
\begin{equation}
A^\prime=BRF(t),
\label{CI2}
\end{equation}
where $f(r)$ and $F(t)$ are arbitrary functions of integration.

Then taking the $r$- derivative of (\ref{CI1}) and the $t$- derivative of (\ref{CI2}) we obtain

\begin{equation}
\frac{\dot B^\prime}{B}=RFf+\frac{\dot B f^\prime}{Bf},
\label{CI1p}
\end{equation}
\begin{equation}
\frac{\dot A^\prime}{A}=RFf+\frac{A^\prime \dot F}{AF}.
\label{CI2p}
\end{equation}

Combining the two equations above we get 
\begin{equation}
\frac{\dot A^\prime}{A}-\frac{\dot B^\prime}{B}=\frac{A^\prime \dot F}{AF}-\frac{\dot B f^\prime}{Bf},
\label{CI3}
\end{equation}
whose  solution is
\begin{equation}
A=C\frac{BF}{f},
\label{CI4}
\end{equation}
where $C$ is a constant of integration with units of $[length]$.

Feeding back (\ref{CI4}) into (\ref{CI1}) it follows that
\begin{equation}
\frac{\dot B}{B}=CF \Longrightarrow B=B_1(t)B_2(r),
\label{CI4a}
\end{equation}
consequently (using (\ref{CI1})), $A$ is also a separable function implying (by performing a reparametrization of $t$) that  $A=A(r)$.

Taking the $t$-derivative of (\ref{CI4}) we get
\begin{equation}
\frac{\dot F}{F}=-\frac{\dot B}{B}.
\label{CI4b}
\end{equation}
On the other hand from (\ref{CI1}) and (\ref{CI4}) we have
\begin{equation}
\frac{\dot B}{B}=CF.
\label{CI4bc}
\end{equation}
Feeding the above expression into (\ref{CI4b}) and integrating we obtain

\begin{equation}
F=\frac{1}{Ct+\beta},
\label{CI5}
\end{equation}
where $\beta$ is a constant of integration with units $[length^2]$.

Feeding back (\ref{CI4}) into (\ref{CI2}) we obtain
\begin{equation}
\frac{A^\prime}{A}=\frac{Rf}{C}.
\label{CI6}
\end{equation}
Also,  using (\ref{CI4}) and (\ref{CI5}) we get
\begin{equation}
B=\frac{A(Ct+\beta)f}{C},
\label{CI7}
\end{equation}
while  from (\ref{U0sigthe}), (\ref{CI4}) and (\ref{CI4bc}), the shear $\sigma$ can be written as
\begin{equation}
\sigma=\frac{f}{B}.
\label{CIs}
\end{equation}

From all the above expressions we obtain for the physical variables
\begin{equation}
8\pi \mu=-\frac{\sigma^2}{f^2}\left[\frac{2R^{\prime\prime}}{R}+\frac{R^{\prime2}}{R^2}-\frac{2R^\prime}{R}\left(\frac{Rf}{C}+\frac{f^\prime}{f}\right)\right]+\frac{1}{R^2},
\label{CImu}
\end{equation}
\begin{equation}
4\pi q=-\frac{\sigma^2}{f}\frac{R^\prime}{R},
\label{CIq}
\end{equation}
\begin{equation}
8\pi P_r=\frac{\sigma^2}{f^2}\left( \frac{2fR^\prime}{C}+\frac{R^{\prime2}}{R^2}\right)-\frac{1}{R^2},
\label{CIPr}
\end{equation}

\begin{equation}
8\pi P_\bot=\frac{\sigma^2}{f^2}\left(\frac{R^\prime f}{C}+\frac{R^{\prime\prime}}{R}-\frac{R^\prime f^\prime}{Rf}\right).
\label{CIPt}
\end{equation}
\noindent In order to obtain a simple specific model, let us assume for $A$  the form
\begin{equation}\label{eA}
  A=bR^n\quad \Rightarrow\quad f=\frac{nCR^\prime}{R^2},
\end{equation}
where $b$, $n$ and $C$ are arbitrary constants.

\noindent Then, the  physical variables read

\begin{eqnarray}
  8\pi \mu &=& \frac{\sigma^2 R^2}{n^2 C^2}(2n-5)+\frac{1}{R^2}, \\
  8\pi P_r&=& \frac{\sigma^2 R^2}{n^2 C^2}(2n+1)-\frac{1}{R^2}, \\
  8\pi P_\bot &=& \frac{\sigma^2 R^2}{n^2 C^2}(n+2).\\
  4\pi q &=& -\frac{\sigma^2 R}{n C}.\label{qu0}
\end{eqnarray}

\noindent  Although in general this model does not require the existence of a Minkowskian cavity surrounding the center, if we assume that it exists then it follows from (\ref{j3in}) and  (\ref{CIq})  that $\sigma=0$, producing a non--dissipative model.

On the other hand  (\ref{CIq}) - (\ref{CIPr})  together with (\ref{20lum})  imply that

  \begin{equation}\label{SiCI}
    \sigma=\frac{nC}{R^2}\frac{1}{\sqrt{2n+1+\frac{2nC}{R}}}\quad \Rightarrow \quad \dot{\sigma}=0\quad \Rightarrow \quad C=0,
  \end{equation}
  i.e the solution is static. Thus this model does not satisfy Darmois conditions.
Finally, from (\ref{V2}), (\ref{eA})  and (\ref{qu0}) we obtain for the temperature of this model
\begin{equation}
  T(t,r) = \frac{\tau C^2}{2\pi b^2n\kappa R^{2n}(Ct+\beta)^2}+\frac{C}{4\pi b \kappa (C t+\beta)}\frac{\ln R}{ R^n}+T_0(t).\label{TemCI}
\end{equation}

\section{CONCLUSIONS}
We have studied in detail the consequences emerging from the vanishing complexity factor condition plus the quasi--homologous evolution. To obtain specific models we have introduced further, different conditions on the kinematical variables defined in Section 6.
It has been shown that one such condition is particularly suitable for describing the evolution
of a fluid distribution endowed with a cavity surrounding the center. 

All equations governing
the dynamics under the   conditions considered here 
have been written down and several models have been
presented. Some of them satisfy Darmois conditions on both delimiting hypersurfaces, precluding thereby the appearance of shells on either of these hypersurfaces.  Other
models result from relaxing Darmois conditions and adopting Israel junction conditions across
shells.

We have considered nondissipative as well as dissipative systems. In the former case it was shown that  by replacing the homologous condition assumed in \cite{2} by the quasi--homologous condition defined in section 4, we were able to obtain a great deal of models satisfying the vanishing complexity factor, in contrast with the unique model existing under the homologous condition.

 In the dissipative case we used a transport equation derived from a causal theory of dissipation, which allowed us to calculate the explicit expressions of the temperature for each model. The interest of these expressions resides in the fact that they contain two type of contributions; on  the one hand contributions from the terms proportional to the relaxation time. These terms are related to  the transient processes  occurring  before relaxation;  they play a fundamental role for time scales of the order of  (or smaller than)  the relaxation time, but of course their contribution remain valid  for all time scales.  On the other hand there are the contributions from terms that do not contain $\tau$,  these are associated to the stationary  dissipative regime. Thus the expressions obtained for the temperature encompass all the thermal history of the compact object, including the epoch before relaxation.

Two main issues motivated the present work. On the one hand we wanted to bring out general physical properties inherent to all dissipative models satisfying conditions  (\ref{ch1})  and  $Y_{TF}=0$. With respect to this  question our results are not particularly encouraging since we were unable to detect any distinct physical behaviour  characterizing all models.

One the other hand we wanted to use (\ref{ch1})  and  $Y_{TF}=0$ as heuristic conditions to find exact analytical solutions to Einstein equations describing collapsing dissipative fluid spheres, and which could be used eventually to model  some interesting astrophysical scenarios. In this case the results are much more promising.

Indeed, one possible application of the presented results  is the modeling
of evolution of cosmic voids.  These are underdensity
regions observed in the large-scale matter distribution in the universe (see \cite{vz,7v,1v,2v,5v,3v,4v,5vu,5vn} and references
therein). In general voids are neither empty nor spherical. However, for simplicity they are usually described as vacuum spherical
cavities surrounded by a fluid, as we do here. 
For cavities with sizes of the order of 20 Mpc or smaller, the assumption of  a  spherically symmetric spacetime outside the cavity is quite reasonable, however for larger cavities, say on scales equal or larger than 150-300 Mpc.,  for which  the observed universe can be considered homogeneous, it should be more appropriate to consider their embedding in an expanding Lema\^{\i}tre-Friedmann-Robertson-Walker spacetime (for the specific case of void modeling in expanding universes see \cite{bill}, \cite{Torres} and references therein).

The relevance of voids in cosmological studies stems from the fact that it seems that  the actual universe
has a spongelike structure, dominated by voids \cite{vn7}.  This picture is supported by  observations  suggesting that about a half 
 of the presented volume of the universe is in voids of a characteristic scale $30h^{-1} Mpc$,
where $h$ is the dimensionless Hubble parameter, $H_0 = 100 h km s^{-1} Mpc^{-1}$  \cite{vn8} . However, voids
of very different scales may be found, from minivoids \cite{3vn} to supervoids \cite{4vn}. 

We would like to stress that our purpose here has not been to generate specific models of any observed
void, but rather to call the attention to the potential of the purely areal evolution condition for
such a modeling, providing all necessary equations for their description. Models of voids within the thin wall approximation have also been considered in \cite{v4,v5,v6}.

Finally, let us mention two additional  possible astrophysical  applications of our results:
\begin{itemize}

\item Possibly, some of our solutions could be used as toy models of localized systems such as
supernova explosions. It is worth stressing  that for these scenarios, the Kelvin--
Helmholtz phase is of the greatest relevance \cite{b}. 

\item Also, as mentioned in the Introduction the homologous condition appears to be too stringent, since in the non--dissipative case it leads to a unique model. Indeed, for this latter case, the homologous condition implies $Y_{TF} =0$ and produces the simplest configuration (Friedman--Robertson--Walker), which is the only one evolving homologously and satisfying $Y_{T F} = 0$. Therefore our approach which implies a relaxing of the homologous condition, could lead to more sophisticated models of the Universe, as for example the one described in \cite{apj}.
 \end{itemize}
 \section{Acknowledgments}
This work was partially
supported by Ministerio de Ciencia, Innovacion y Universidades. Grant number: PGC2018--096038--B--I00, and  Junta de Castilla y Leon. 
 Grant number:  SA083P17.

\end{document}